# X-ray Diffraction Analysis of $Cu^{2+}$ Doped $Zn_{1-x}Cu_xFe_2O_4$ Spinel Nanoparticles using Williamson-Hall Plot Method

Dinesh Kumar[1,a)], Abhishek Kumar[2,3,b)], Rajiv Prakash[1,c)] and Akhilesh Kumar Singh[1,d)*]

[1]*School of Materials Science and Technology, Indian Institute of Technology (Banaras Hindu University), Varanasi-221005, Uttar Pradesh, India*
[2]*Central Instrumental Facility Centre, Indian Institute of Technology (Banaras Hindu University), Varanasi-221005, Uttar Pradesh, India*
[3]*Acharya Narendra Dev College, University of Delhi, New Delhi-110019, India*

Email: [a)]dineshiitbhu@gmail.com, [b)]abhiphy2015@gmail.com, [c)]rprakash.mst@iitbhu.ac.in, *[d)]aksingh.mst@iitbhu.ac.in

**Abstract**. The nanoparticles (NPs) of $Zn_{1-x}Cu_xFe_2O_4$ (ZCFO) spinels with x = 0, 0.2, 0.4, 0.6 and 0.8 were synthesized by a sol-gel combustion method using acetate precursor. The synthesized NPs of ZCFO were characterized by X-ray diffraction (XRD) analysis using Rietveld structure refinement. The Rietveld refinement of the XRD patterns revealed that the ZCFO spinels crystallize into diamond cubic structure with $Fd\bar{3}m$ space group. The lattice constant and unit cell volume for ZCFO NPs shrink with enhancing doping concentration of $Cu^{2+}$ ion. The crystalline growth in the NPs of ZCFO was examined by peak broadening present in the XRD pattern. The Williamson-Hall (W-H) plot method was used to study the individual role of crystallite sizes and lattice strain on the peak broadening of the NPs of ZCFO spinels. High-resolution scanning electron microscopy was also done to confirm the particle size.

## 1. INTRODUCTION

Among the various ferrites, A-site substituted spinel ferrites ($AFe_2O_4$, where A is divalent cation occupies tetrahedral site and trivalent Fe occupies octahedral site) have attracted more attention of the researchers due to their technological and scientific applications in electronic, magnetic and microwave devices [1-2]. The properties of the spinel ferrites depend mostly on the type of substituting metal ions concentration and their distribution over tetrahedral and octahedral sites [2]. Cu-substituted spinel ferrites are promising candidates in various applications. Substitution of copper into spinel ferrites causes considerable variations in their properties, mainly structural ones, as it may probably generate a lattice distortion due to Jahn-Teller effect and also affect magnetic and dielectric properties. It also enhances the catalytic performance of the ferrite with substitution of Cu, due to the different redox properties of Cu [3]. Recent investigations of spinel ferrites indicated that Zinc ferrites in the nanometric samples show anomalous magnetic properties [1]. There are several studies of the electrical, optical and magnetic behaviors of Cu-substituted Zn ferrites [1-3]. These studies manifested that with increasing Cu concentration, the magnetic properties are significantly changed from superparamagnetic to ferromagnetic, the energy band gap reduces, and a higher dielectric constant is observed [1]. $ZnFe_2O_4$ ferrite is a normal spinel i.e. divalent Zn occupies the position of tetrahedral void and trivalent Fe occupies octahedral void, showing paramagnetic state at room temperature. Various methods were used to synthesize the nanometric samples of ferrites, among which sol-gel method is easy, economical and simple to process [4].

In this article, we report the composition dependent structural changes in the Cu-substituted $Zn_{1-x}Cu_xFe_2O_4$ with x = 0.0, 0.2, 0.4, 0.6 and 0.8 ferrites. Structural analysis was carried out by Rietveld structure refinement using X-ray diffraction data. Crystallite size and lattice strain were studied with the help of Williamson-Hall (W-H) plot method. Microstructure of the samples was studied by HRSEM micrographs.

## 2. SYNTHESIS AND EXPERIMENTAL DETAILS

The nanocrystalline samples of $Zn_{1-x}Cu_xFe_2O_4$ spinel ferrites were prepared using stoichiometric amounts of $Zn(CH_3COO)_2$ (99.99%, Sigma-Aldrich), $Cu(CH_3COO)_2$ (99.99%, Sigma-Aldrich) and $Fe(CH_3COO)_2$ (99.99%, Sigma-Aldrich) by sol-gel combustion method. Glycine (99.5%, Hi-media) was used as fuel for the combustion. The stoichiometric amount of starting materials were dissolved in de-ionized water separately and mixed together in a larger beaker under continuous stirring on a magnetic hot plate at 200°C. After 4-5hrs of continuous stirring the solvent evaporates and mixed precursor solution converted into gel. The obtained gel was dried at same temperature and dried powder was collected and kept at 300°C in a muffle furnace for the complete combustion. Large amount of various gases produced during combustion process. The obtained powder after combustion of the sample was calcined at 600°C for 8hrs to get pure phase of ZCFO NPs. The crystallinity and the phase purity of the samples were examined by XRD. The XRD data were collected in the 2θ range 10°-100° using Rigaku Miniflex600, X-ray diffractometer with hybrid pixel array detector. The Rietveld crystal structure refinement was done by using FullProf Suite [5]. The crystallite size and lattice strain were estimated by using Williamson-Hall plot method. The microstructure of the sample was studied by high-resolution scanning electron microscope (HRSEM, FEI, Nova NanoSEM 450).

## 3. RESULTS AND DISCUSSION

### 3.1. Crystal Structure

To investigate the crystal structure of the nanocrystalline samples of the pure and copper doped zinc ferrites, $Zn_{1-x}Cu_xFe_2O_4$ (x = 0.0, 0.2, 0.4, 0.6 and 0.8), we show in **Fig.1** the room temperature XRD profiles in the two theta range 10°-100°. The XRD patterns display good crystalline nature of the samples. No any major impure secondary phase was detected from the XRD patterns. The XRD patterns shown in **Fig. 1** can be indexed using the structure of standard cubic $ZnFe_2O_4$ spinel ferrite ICDD File No. 89-7412. On doping of $Cu^{2+}$ ion at Zn-site in $Zn_{1-x}Cu_xFe_2O_4$, there is no structural transition to any other phase. Hence, all the samples crystallize into face-centered cubic structure with space group $Fd\bar{3}m$. The inset of **Fig. 1** shows maximum intense Bragg's reflection (311) in the two theta range 34.5°-36.0°, which shifted towards higher angle side with increasing doping concentration of $Cu^{2+}$ ion. This shows that lattice constant and unit cell volume are decreasing as the concentration of dopant ($Cu^{2+}$ ion) is increased in $Zn_{1-x}Cu_xFe_2O_4$ ferrites.

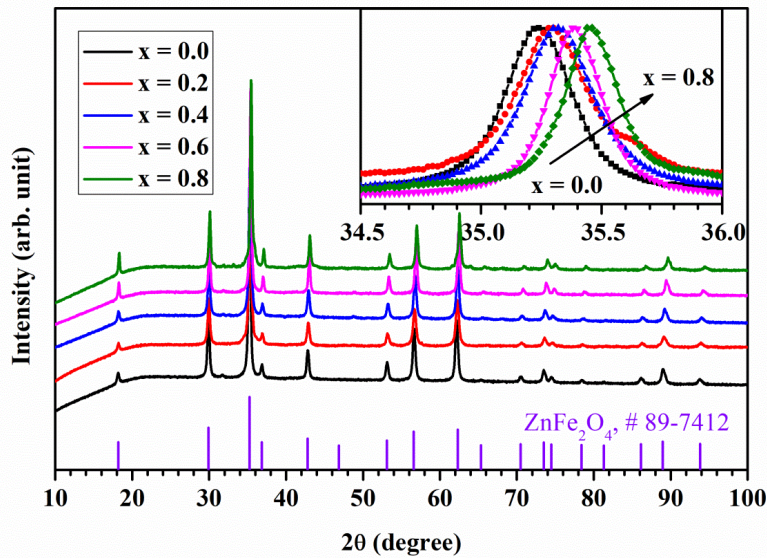

**Fig. 1**: Room temperature X-ray diffraction patterns for $Zn_{1-x}Cu_xFe_2O_4$ spinel ferrite (x = 0.0, 0.2, 0.4, 0.6 and 0.8) with standard pattern of $ZnFe_2O_4$ ferrite. The inset shows selected Bragg's peak between 34.5°-36.0°.

Cubic $Fd\bar{3}m$ space group was used for Rietveld structure refinement of the ZCFO NPs. In the process of Rietveld refinement, we considered that divalent $Cu^{2+}$ ion replaces $Zn^{2+}$ ion. In the unit cell of cubic $Fd\bar{3}m$ space group for ZCFO, $Zn^{2+}/Cu^{2+}$, $Fe^{3+}$ and $O^{2-}$ ions were considered to occupy the sites 8$b$ (3/8, 3/8, 3/8), 16$c$ (0, 0, 0) and 32$e$ ($\delta x$, $\delta y$, $\delta z$), respectively [6]. The atomic/ionic coordinates obtained after Rietveld structure refinement are given in **Table 1**. The acceptable value of $\chi^2$ and very good fit between the experimentally observed and Rietveld calculated XRD patterns are obtained. **Figs. 2(a-b)** demonstrate Rietveld fits between experimental (solid dots) and calculated (continuous curve) XRD patterns for ZCFO ferrites with x = 0.0 and x = 0.4, respectively. The Rietveld refined lattice constant "a" and unit cell volume "V" decrease monotonically from a = 8.4418(1) Å to 8.3953(2) Å and V = 601.60(1) Å$^3$ to 591.70(3) Å$^3$ with increasing $Cu^{2+}$ ion concentration from x = 0 to x = 0.8, respectively. The reduction in the lattice constant or unit cell volume is due to replacement of bigger ionic sized $Zn^{2+}$ ions with ionic radius $r_{Zn}^{2+}$ = 0.60 Å by smaller ionic sized $Cu^{2+}$ ions $r_{Cu}^{2+}$ = 0.57 Å [7-8]. The refined values of lattice constant and unit cell volume for ZCFO NPs are in good agreement with the values reported by earlier authors [1-2]. The values of lattice constant, unit cell volume and $\chi^2$ for various compositions are tabulated in **Table 1**. **Fig. 2(c)** displays continuously decreasing lattice constant and unit cell volume as a function of doping concentration of $Cu^{2+}$ ion. **Fig. 2(d)** shows ball and stick model for $ZnFe_2O_4$ spinel (ZFO) drawn using crystallographic information file (.cif) with the help of VESTA software. The ball and stick model clearly shows that $Zn^{2+}$ ions occupy tetrahedral voids, while $Fe^{3+}$ ions occupy octahedral voids present in the unit cell. Hence, the unit cell of ZFO contains tetrahedra of $ZnO_4$ and octahedra of $FeO_6$ (see **Fig. 2(d)**).

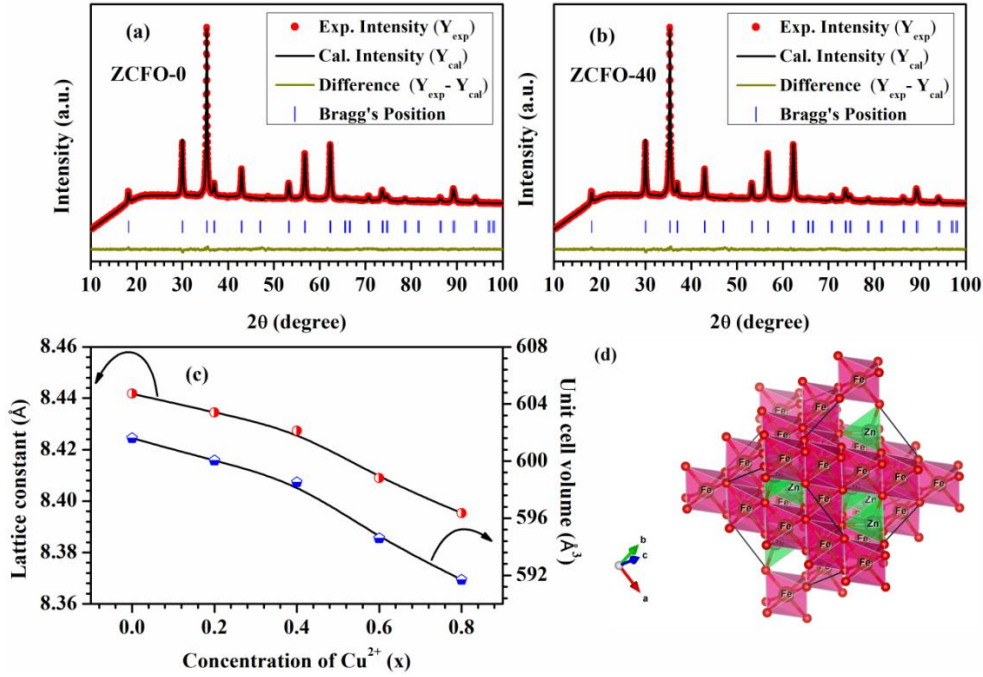

**Fig. 2**: Rietveld fits for $Zn_{1-x}Cu_xFe_2O_4$ spinel ferrite (a) x = 0.0 (ZCFO-0) and (b) x = 0.4 (ZCFO-40). (c) Variation of lattice constant (a) and unit cell volume with doping concentration of $Cu^{2+}$ ion. (d) Ball and Stick model for $ZnFe_2O_4$ ferrite.

### 3.2. Estimation of crystallite size and lattice strain

The XRD can be used to estimate peak broadening along with crystallite size and lattice strain appearing due to defects [10]. The crystallite size and lattice strain of the NPs of ZCFO ferrites were calculated using modified Scherrer's formula and Williamson-Hall (W-H) method. According to W-H method broadening in the peak due to crystallite size ($\beta_d$) and lattice strain ($\beta_s$) can be given by the equation:

$$\beta_{hkl} = \beta_d + \beta_s$$
$$\beta_{hkl} = k\,\lambda/(d\cos\theta_{hkl}) + 4\varepsilon\,\tan\theta_{hkl}$$

where, λ is the wavelength of $Cu_{K\alpha}$ X-ray radiation (λ = 1.5406 Å), d is average value of crystallite size, k is a constant equal to 0.89, $\theta_{hkl}$ is Bragg's angle, $\beta_{hkl}$ is the full width at half-maximum corresponding to (hkl) Bragg's peak, and ε is the average value of lattice strain [9-11]. Rearranging the above equation gives:

$$\beta_{hkl} \cos\theta_{hkl} = k \lambda/d + 4\varepsilon \sin\theta_{hkl}$$

This equation represents the uniform deformation model (UDM) or equation for W-H plot, where the strain was supposed to be uniform in all the crystallographic directions. The W-H plots between (β cosθ) and (4 sinθ) were employed to estimate crystallite size and lattice strain. Accordingly, the y-intercept and the slope of the fitted straight line give crystallite size and lattice strain, respectively.

**TABLE 1**. Structural parameters (a=b=c and V), goodness of fit ($\chi^2$), crystallite size (d), lattice strain (ε), average bond lengths ($<d_{Zn-O}>$ & $<d_{Fe-O}>$) and atomic coordinates of O-atom for $Zn_{1-x}Cu_xFe_2O_4$ spinels obtained after Rietveld structure refinement.

| Parameters | x = 0.0 | x = 0.2 | x = 0.4 | x = 0.6 | x = 0.8 |
|---|---|---|---|---|---|
| a=b=c (Å) | 8.4418(1) | 8.4345(2) | 8.4273(1) | 8.4089(1) | 8.3953(2) |
| V (Å³) | 601.60(1) | 600.04(3) | 598.51(2) | 594.59(2) | 591.70(3) |
| $\chi^2$ | 1.02 | 1.91 | 1.07 | 1.87 | 1.94 |
| O (δx = δy = δz) | 0.2414(1) | 0.2434(2) | 0.2436(1) | 0.2436(2) | 0.2409(3) |
| $d_{Zn-O}$ (Å) | 1.9534(10) | 1.9218(17) | 1.9205(14) | 1.9141(14) | 1.950(3) |
| $d_{Fe-O}$ (Å) | 2.0404(10) | 2.0549(16) | 2.0530(14) | 2.0497(14) | 2.025(3) |
| $d_{Zn-Fe}$ (Å) | 3.49978(5) | 3.49676(7) | 3.49381(6) | 3.5036(14) | 3.48051(9) |
| d (nm) | 26.8 | 31.7 | 33.0 | 37.9 | 41.3 |
| ε | $-3.76\times10^{-5}$ | $7.70\times10^{-4}$ | $5.44\times10^{-4}$ | $1.50\times10^{-4}$ | $2.41\times10^{-4}$ |

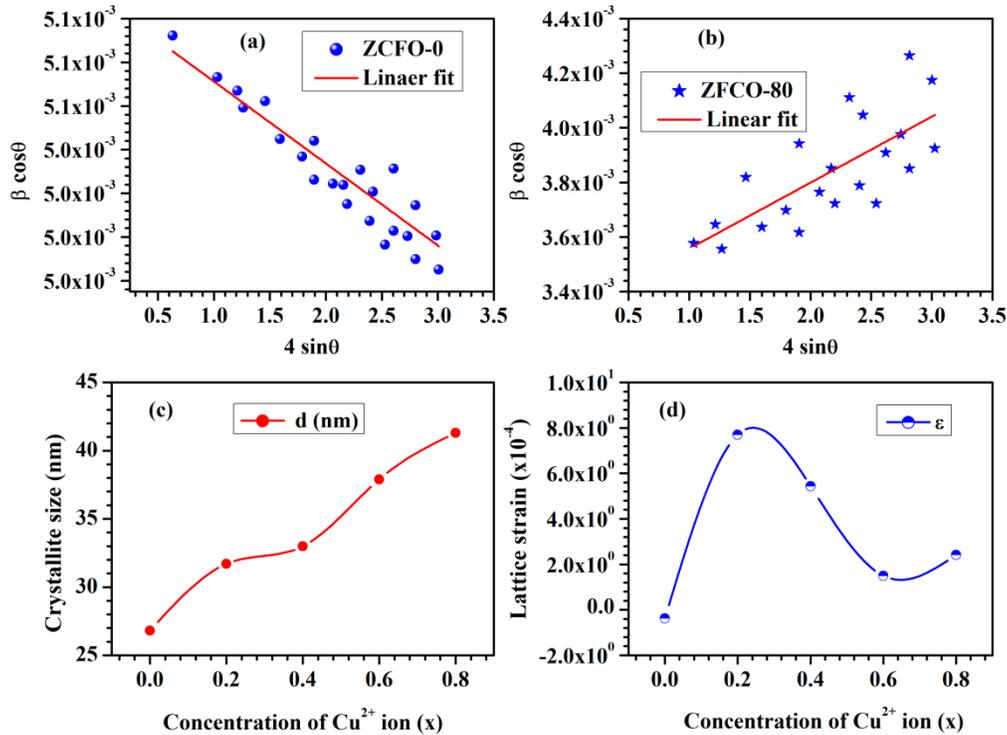

**Fig. 3**: Williamson-Hall plots for $Zn_{1-x}Cu_xFe_2O_4$ spinel ferrites (a) x = 0.0 (ZCFO-0) and (b) x = 0.8 (ZCFO-80). Evolution in (c) crystallite size and (d) lattice strain as a function of doping concentration of $Cu^{2+}$ ion for $Zn_{1-x}Cu_xFe_2O_4$ ferrites.

**Figs. 3(a-b)** represent W-H plots for ZCFO ferrites with x = 0.0 and 0.80, respectively. The average value of the crystallite size increases with increasing doping concentration of $Cu^{2+}$ ions and are found to be 26.8, 31.7, 33.0, 37.9 and 41.3 nm for ZCFO ferrites with x = 0.0, 0.2, 0.4, 0.6 and 0.8, respectively. The enhancement in the crystallite size decreases peaks broadening in the XRD patterns. The W-H plot for the NPs of the sample with x = 0.0 showed

negative strain, which may be due to lattice shrinkage. Doping of $Cu^{2+}$ ion at Zn-site introduces a positive strain, which decreases with increasing concentration of $Cu^{2+}$ ion (x) due to increase in crystallite size. **Figs. 3(c-d)** show variation of crystallite size and lattice strain with doping concentration of $Cu^{2+}$ ion (x). The values of the crystallite size and lattice strain are given in **Table 1**.

**Fig. 4(a)** shows high-resolution scanning electron microscopy image of ZCFO spinel ferrite for x = 0.40. The average particle size of the sample falls into the range 30-45 nm, which confirms that the synthesized sample is nanocrystalline. The average value of the particle size was estimated considering as many particles as possible from the micrographs using ImageJ software and found to be ~ 40 nm [11]. A histogram for the particle size was plotted to find out the most probable value of particle size by fitting Gaussian curve. The distribution of particle size in form of histogram is shown in **Fig. 4(b)**. Gaussian fit gives the most probable value of particle size ~ 37.6 nm.

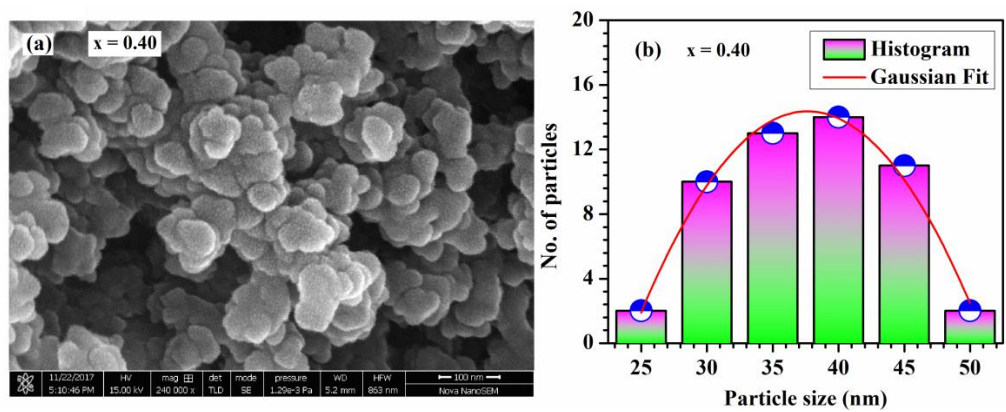

**Fig. 4**: (a) HRSEM micrograph for ZCFO ferrite with x = 0.40. (b) Histogram for the distribution of particle size along with Gaussian fit.

## 4. CONCLUSIONS

The phase pure NPs of $Zn_{1-x}Cu_xFe_2O_4$ ferrites with x = 0.0, 0.2, 0.4, 0.6 and 0.8 were synthesized by sol-gel combustion method and characterized by powder X-ray diffraction. The Rietveld analysis of the XRD patterns shows that the NPs of ZCFO ferrites crystallize into cubic structure with space group $Fd\bar{3}m$. The lattice constant and unit cell volume for ZCFO NPs decreases with enhancing doping concentration of $Cu^{2+}$ ion, due to smaller size of $Cu^{2+}$ ions replacing the $Zn^{2+}$ ions. The crystalline growth in the NPs of ZCFO was scrutinized by studying peak broadening in the XRD using W-H plot method. The calculated values of crystallite size increase with increasing concentration of $Cu^{2+}$ ions and reduce the peak broadening. The analysis of the XRD patterns and HRSEM micrograph shows that synthesized samples are nanocrystalline. The lattice strain for the pure sample i.e. $Zn_{1-x}Cu_xFe_2O_4$ with x = 0 shows negative value which indicates compressive nature of the strain. However, the values of lattice strain for doped samples with x ≠ 0 show positive value of tensile nature and decreases with increase in doping concentration of $Cu^{2+}$ ions and crystallite size.